\documentclass{article}
\usepackage{emulateapj}

\input{epsf}

\def\t0{t_0}

\newcommand{\bez}{\begin{eqnarray*}}
\newcommand{\eez}{\end{eqnarray*}}
\newcommand{\be}{\begin{equation}}
\newcommand{\ee}{\end{equation}}
\newcommand{\beq}{\begin{eqnarray}}
\newcommand{\eeq}{\end{eqnarray}}
\newcommand{\bc}{\begin{center}}
\newcommand{\ec}{\end{center}}

\begin{document}

\title{
Evidence for a Fast Decline in the Progenitor Population of Gamma Ray 
Bursts  and the Nature of their Origin}

\author{B. E. Stern\altaffilmark{1,2,3}, 
J-L. Atteia\altaffilmark{4}, 
K.Hurley\altaffilmark{5}}

\altaffiltext{1}{Institute for Nuclear Research, Russian Academy of Sciences,
Moscow 117312, Russia}
\altaffiltext{2}{Astro Space Center of Lebedev Physical Institute,
Moscow, Profsoyuznaya 84/32, 117810, Russia}
\altaffiltext{3}{SCFAB, Stockholm Observatory, SE-106 91 Stockholm, Sweden}
\altaffiltext{4}{CESR, 9, avenue du Colonel Roche, 31029 Toulouse Cedex, France}
\altaffiltext{5}{University of California Space Sciences Laboratory, Berkeley, CA 94720-7450}

\begin{abstract}
We show that the source population of long gamma-ray bursts (GRBs) has 
declined by at least a factor of 12 (at the 90\% confidence level) since 
the early stages of the Universe ($z \sim 2 - 3$). This result 
has been obtained using  the
combined BATSE and \it Ulysses \rm GRB brightness distribution and the 
detection of four GRBs  with known redshifts brighter than
10$^{52}$ erg s$^{-1}$ in the 50 - 300 keV range at their peak. 
The data indicate that the decline of the GRB source population is as fast as, or even faster than, the 
measured decline of the star formation rate.    
Models for the evolution of neutron star binaries predict a significantly 
larger number of apparently bright GRBs than observed. 
Thus our results give independent support to the hypernova model,
which naturally explains the fast decline in the progenitor population.
 
\end{abstract}
\keywords{gamma-rays: bursts -- methods: data analysis} 


\section{Introduction}

The cosmological evolution of GRB progenitors 
at redshifts $z<2$ can, in 
principle, reveal their nature. Indeed, we have unambiguous star formation 
data (hereafter SF; see Porciani \& Madau 2001 and references therein) for the 
declining stage which 
started after $z\sim2$, which we can use 
as a reference evolutionary curve. If GRB progenitors follow this 
curve or decline even faster than it, then we have to conclude that 
GRBs are most probably associated with the collapse of supermassive stars 
(hypernovae, or ``failed supernovae'' as originally suggested by Woosley 1993;
see also Paczynski 1998 and MacFadyen \& Woosley 1999). If the decline of GRBs is slower 
than the SF decrease then the  coalescing neutron star binary model would be
supported, as it naturally provides a delay between star formation 
and bursts.

 GRB afterglow observations provide three lines of evidence in favor of the hypernova
model as an explaination for the long duration GRBs. First, a large fraction of
the afterglows are found near the 
central regions of their host galaxies (Bloom, Kulkarni \& Djorgovski 2001).  Second, features have
been found in the light curves of three afterglows which can be
interpreted as a supernova component (e.g. Bloom et al. 1999, Lazzati et al. 2001).  And third,
absorption features in some x-ray afterglow spectra (Galama \& Wijers, 2001)
and an emission iron K$\alpha$ line  (Piro et al., 2000)   indicate a high 
metal column density along the line of sight.  For more details see 
the review of Meszaros (2001) and references therein.
Although none of these facts constitutes a decisive argument by itself, together they strongly
favor the hypernova model.  However, the cosmological evolution of 
GRB sources can provide a new and completely independent test for the
nature of GRB progenitors.

 The problem of deriving the GRB source evolution from the data is not
simple and cannot be solved by a straightforvard cosmological fit to the 
log N - log P distribution with an unknown GRB luminosity function. 
Despite the wealth of statistics on GRBs accumulated by the Burst and Transient
Source Experiment aboard the \it Compton Gamma-Ray Observatory \rm (BATSE)
(see Fishman et al. 1989),   
the bright end of the distribution still contains too few events to provide 
a conclusive $\chi^2$ fit. For a review of cosmological fits to the 
log N - log P distribution see, e.g.,  Bulik (1999).

 Stern, Tikhomirova \& Svensson (2001 - 
hereafter STS), 
using the redshift data and the BATSE GRB sample, demonstrated 
the cosmological decline of GRB progenitors. A fast decline, similar to that of SF or even
faster, was preferred by the data, but the 
statistics were insufficient to distinguish between the predictions of the hypernova 
and neutron star (NS) binary models at a significant level.

 In this work we incorporate the data of the \it Ulysses \rm GRB experiment, which more than double
the number of strong GRBs and allow a more reliable 
interpretation of the redshift data. The main objective is to achieve
a scientifically meaningful constraint on the NS binary model.   

In \S 2 we describe the data and the procedure used to cross- 
calibrate the \it Ulysses \rm and BATSE GRB samples. In \S3 we outline the model
fitting procedure, including the cosmological model, the parameterized 
luminosity function, and the  GRB source evolution hypothesis.
In \S 4 we present our results and show that the data require a very fast
GRB progenitor decline, seriously challenging existing NS binary models.  


\section{The data}


 We have used three independent data sets.
The first contains 3255 
 BATSE GRBs with durations longer than 1 s, found by 
Stern et al.  (2000, 2001) in the off-line scan of the 1.024 s time resolution BATSE continuous 
daily records for the entire 9.1 yr BATSE 
mission \footnote[6]
{see http://www.astro.su.se/groups/head/grb\_archive.html}. 
This is the largest essentially uniform GRB sample, and its 
efficiency matrix has been measured using a test burst method. 
 The second data set is the \it Ulysses \rm sample,  
consisting of only bright GRBs,
which are the most important ones for the aim of the present work. The
\it Ulysses \rm GRB detector has amassed well over 10 years of data to date,
and since the detector is in interplanetary space and is neither Earth- nor spacecraft-occulted, it has $\approx 4 \pi$ sr sky exposure and a larger 
effective duty cycle than BATSE (useful data are recovered for more than 95\% of the mission), thus more then doubling the number
of bright GRBs.  The \it Ulysses \rm GRB data on over 800 bursts have appeared in eight
catalogs so far (Hurley et al. 1999a,b; Laros et al. 1997, 1998;
Hurley et al., 2000a,b,c; Hurley et al. 2001a); the instrument description may be
found in Hurley et al. (1992).
 The third data set consists of the GRB redshift data, or more specifically the data on the four 
intrinsically brightest events out of 23 GRBs with measured redshifts (up to 
November 2001)\footnote[7]
{see, e.g., http://www.aip.de/~jcg/grb.html} .

 The first two data sets were combined to form a single log N - log P  
distribution, i.e. the number of events versus the {\it apparent} peak 
brightness, $P_a$, while the third data set was used to constrain the bright 
end of the hypothetical {\it intrinsic} peak brightness ($P_i$) distribution 
(the luminosity function).

\subsection{The BATSE and {\it Ulysses} samples, their cross-calibration, and the 
joint log N - log P distribution} 

 The BATSE sample includes 3255 ``long'' (duration $> 1$ s) GRBs selected from 3906 GRBs 
in the sample of Stern et al. (2001), with the requirement that the counts in 
the second highest time bin exceed 50\% of the counts in the highest time bin.
This sample corresponds to 9.1 years of 
GRO observations with an average exposure factor 0.47. The latter includes 
both Earth-blocking and the average duty cycle for
useful 1.024 s continuous records.

The \it Ulysses \rm sample covers 10.3 years (Dec 1990 - Feb 2001) which overlap
the entire BATSE mission. The average exposure factor of the GRB experiment is $\sim 0.96$.
This includes data outages, as well as periods when the background was high due to solar
proton events.  The short 
events were removed from the sample using the same criterion that we applied to BATSE 
GRBs.
The effective energy range of the \it Ulysses \rm GRB detectors is 
$\approx$ 25 - 150 keV, while the BATSE data used for this study are for the 50 - 300 keV band.

Although both the \it Ulysses \rm GRB detector and BATSE have quasi-isotropic
angular responses, a direct conversion of Ulysses counts to BATSE peak flux is not 
possible for several reasons.  First, the responses of both experiments have a weak angular dependence.  Second, 
they operate in different
spectral bands.  And third, they have different efficiencies as a function of energy.
 The only realistic way to construct a joint log N - log P distribution is 
to do a cross-calibration using joint \it Ulysses \rm /BATSE events. There are 278
such GRBs down to the \it Ulysses \rm cutoff adopted here of 100 counts s$^{-1}$
above background (the average background rate is 480 counts s$^{-1}$). A scatter 
plot of the \it Ulysses \rm count rate versus BATSE flux is shown in figure 1.

\centerline{\epsfxsize=10cm\epsfysize=10cm {\epsfbox{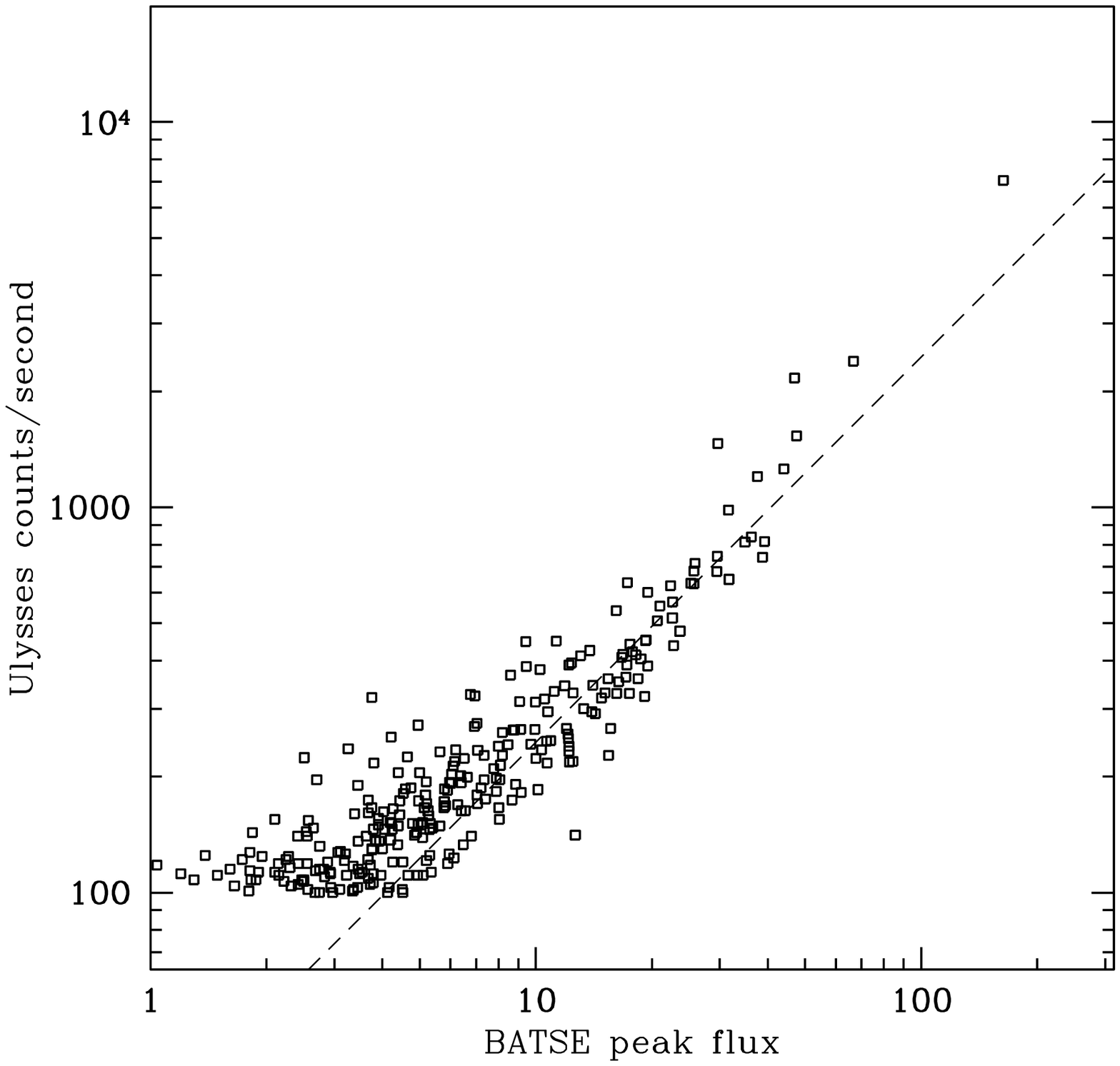}} }
\figcaption{
 Ulysses count rate versus BATSE peak flux for 278 GRBs detected by both experiments.
The sample has been truncated at a \it Ulysses \rm peak count rate of 100 s$^{-1}$.
The dashed line shows the ratio of the \it Ulysses \rm count rate to BATSE flux, 24.6 cm$^2$, which we use as the calibration coefficient.
}
\smallskip

 The tendency towards higher Ulysses/BATSE ratios in the range below 10 
photons cm$^{-2}$s$^{-1}$ is
caused mainly by the Poisson bias (i.e., when one selects the highest Poisson 
fluctuation as a peak) and possibly by the hardness-brightness correlation
in GRBs (see below). At the bright end of the distribution a similar bias 
is evident: 
the \it Ulysses \rm count rates are again systematically higher.
This is probably a saturation effect in the BATSE count rates, resulting from 
slow light emission in NaI scintillator (Meegan \& Preece, 2001).  
This is probably a saturation effect in the BATSE count rates caused by dead time in the large BATSE
detectors.

 To avoid these biases we have restricted the analysis to bursts in the 10 - 40 photons cm$^{-2}$s$^{-1}$ range for the Ulysses/BATSE
calibration. There are 73 events in this range; the average Ulysses/BATSE ratio
is 24.6 cm$^2$, and the rms variance is 5.2. The relatively large variance 
is due to the three reasons cited above.
Note that this flux range 
gives the smallest \it Ulysses \rm /BATSE
ratio and therefore the largest values of \it Ulysses \rm  peak fluxes. Therefore 
it provides a conservative constraint on the decline of the GRB population. 

 In principle there should also be a brightness dependence in the 
\it Ulysses \rm /BATSE ratio caused by the brightness-hardness correlation in GRBs 
(Nemiroff et al., 1994) and the different spectral bands and efficiencies as a function
of energy. This effect may 
contribute to the tendency towards higher \it Ulysses \rm /BATSE ratios for weaker GRBs
(figure 1). We cannot separate this from the Poisson 
bias.  However, 
Atteia (2001) parameterized the correlation between the \it Ulysses \rm count rate and the 50 - 300 keV photon flux as
as $P_a \propto C_u^{1.14}$. 
Thus the brightness dependence is weak and has a negligible effect on the  
results. We prefer to use a constant calibration coefficient because there are insufficient 
data to quantify this dependence more accurately.

 The joint BATSE-\it Ulysses \rm  log N - log P distribution is shown in figure 2.
It includes 77 \it Ulysses \rm  bursts above log$(P_a) = 1.2$. We chose this 
relatively high cutoff to conservatively avoid the Poisson bias in the \it Ulysses \rm  
peak flux estimate. The number of BATSE
GRBs in the range log$(P_a) > 1.2$ is 43 
(the total number of BATSE GRBs in the distribution is 3255).

\centerline{\epsfxsize=10cm\epsfysize=10cm {\epsfbox{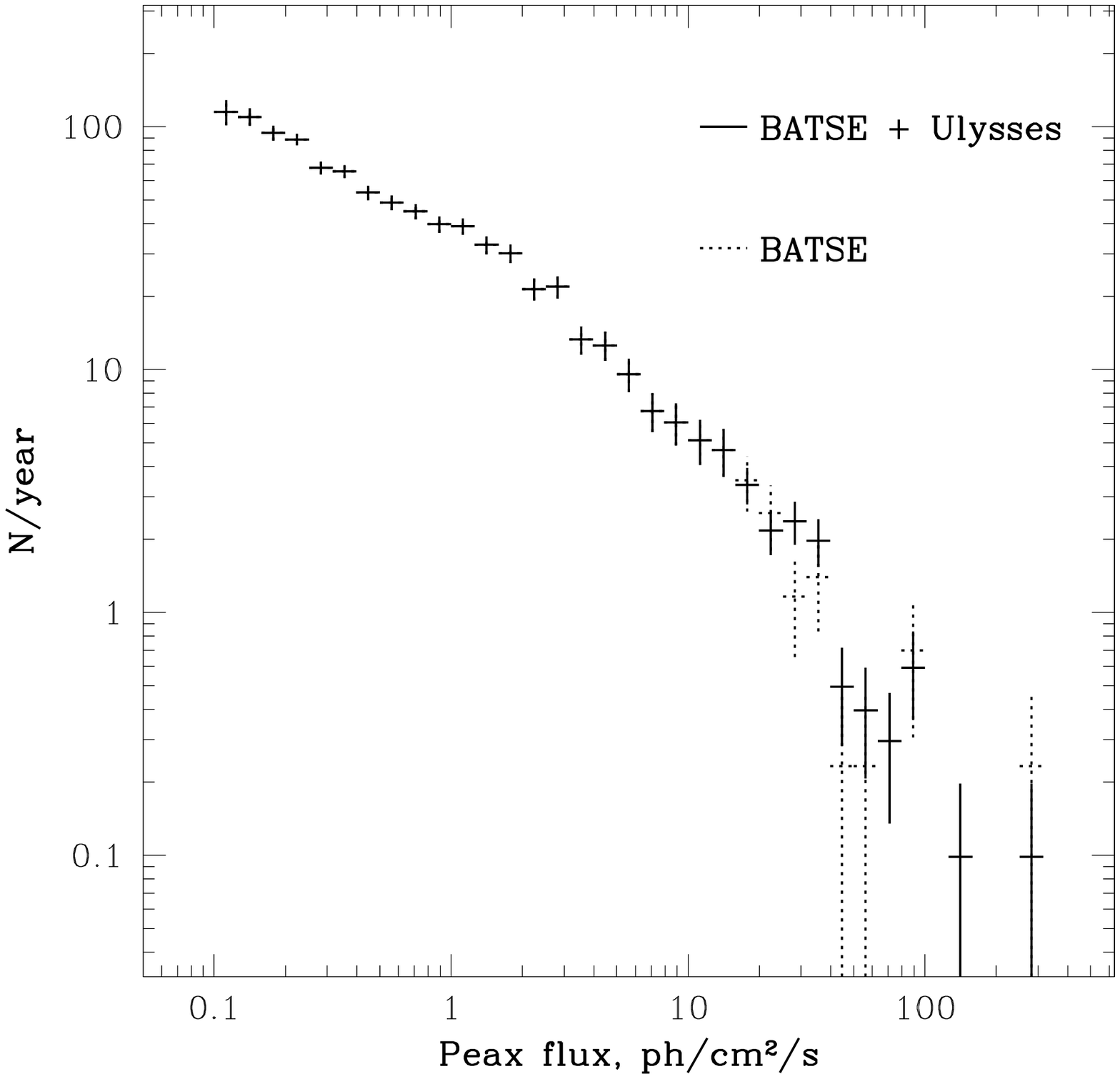}} }
\figcaption{
The joint BATSE - \it Ulysses \rm  log N - log P distribution. 
}
\smallskip

The distribution below log$(P_a)=1.2$ is normalised to the 
total BATSE exposure, 9.1 yr $\times F_b$, where $F_b =0.47$ is the BATSE exposure factor. The normalisation of points above 
log$(P_a)=1.2$ is the sum of the total BATSE/\it Ulysses \rm  exposure during 9.1 yr
and the \it Ulysses \rm -only  exposure during 1.2 yr: 9.1 yr$\times (F_b + F_u - F_b\cdot F_u)
+ $ 1.2 yr$\times F_u$, where $F_u = 0.96$ is the \it Ulysses \rm  exposure factor. 

\subsection{The rate of intrinsically strong GRBs}

The sample of events with known redshift is subject to strong selection
biases and cannot be used directly to determine the luminosity function (see STS).
It does, however, give useful 
information about the existence of intrinsically very bright GRBs.
We can use this fact to constrain the bright end of the hypothetical 
luminosity function: the predicted rate of GRBs with $P_i$ above some threshold at all 
redshifts should correspond to the observed rate. This constraint will 
affect the predicted number of apparently bright GRBs and therefore
constrain the GRB source evolution model.

For convenience  we measure the intrinsic peak brightness $P_i$ as the 
peak flux normalized to $z=1$, taking the K-correction into account.
 The approximate relation between $P_i$ and the isotropic peak 50 - 300 keV
luminosity  is $L = 3\cdot 10^{50}$ erg s$^{-1}$ $P_a$.
STS have chosen the range $P_i > 40$ photons s$^{-1}$ cm$^{-2}$
as a reference for the comparison between the hypothetical luminosity function and the data, and 
we use the same threshold here. 
Four of the intrinsically brigthest bursts are above this threshold:
GRB990123, GRB991216, GRB000131 and GRB010222 , with redshifts of 1.6
 (Djorgovski et al. 1999),
1.02 (Vreeswijk et al. 1999), 4.5 (Andersen et al. 2000), and 1.477  (Stanek et al. 2001)
respectively; their apparent fluxes, $P_a$, in photons s$^{-1}$ cm$^{-2}$, are 
16.4, 67.5, 6.3  (BATSE catalog) and 
22.4  (\it Ulysses \rm  data with our \it Ulysses \rm /BATSE 
calibration) respectively.
Their  $P_i$ values are 45, 69, 84 and 48 photons s$^{-1}$ cm$^{-2}$, respectively.

These 4 events were detected over 4.2 years 
from the beginning of 1997 to March 2001. The latter date corresponds to 
the 
end of the processed \it Ulysses \rm  data that we use here 
(currently there are three GRBs with measured redshifts after March 2001). 

The corresponding rate of GRBs at $P_i > 40$ photons s$^{-1}$ cm$^{-2}$,
$I_{40}$, multiplied by the probability that the burst will be detected and 
localized, its afterglow observed and its redshift measured (hereafter the 
sampling factor, $F_s$)
is 0.95$^{+0.60}_{-0.38}$. Now, in order to estimate $I_{40}$, we have to  
evaluate the sampling factor. 

 The first approach is simply to estimate $F_s$ for \it apparently \rm
bright events,
which would give a reasonable upper limit to the sampling factor for 
\it intrinsically \rm bright events.
 Taking all \it Ulysses \rm  GRBs with peak count rates above 370 s$^{-1}$ 
which corresponds to approximately 15 photons s$^{-1}$ cm$^{-2}$ (62 events
from 1997 January 1 to 2001 March 1) we find redshift data for 4 of them
(two of which are in the above list of four intrinsically bright GRBs).
 Using these numbers we 
estimate the sampling factor as $ F_s \sim 0.064^{+0.038}_{-0.026}$. 
Taking the $1\sigma$ upper limit, 0.1, as a conservative estimate we
obtain the 
rate of intrinsically bright GRBs  $I_{40} \sim 10$.

 The alernative approach is a direct estimate of the efficiency of the 
detectors used for burst localisations. 13 out of 23 redshift measurements 
were done using Beppo-SAX localizations. 7 of the remaining 10 bursts
were localized by the Interplanetary Network (IPN: Hurley et al. 2001b). It is difficult to 
estimate the IPN efficiency, but the BeppoSAX efficiency has a well defined 
upper limit:  the total field-of-view of the two
Wide Field Cameras,
$\sim 0.08$ of the sky\footnote[8]{see http://www.asdc.asi.it/bepposax/}.
Not every localisation, even of a strong burst, is followed by the 
observation of an afterglow and a redshift measurenent. With 
this upper limit on the BeppoSAX efficiency and its share in the redshift 
sample we again obtain an estimate of the sampling factor of $\sim 0.1$.  


 Therefore we adopt the estimate  $I_{40} = 10/$yr as our baseline and, 
to take the poor statistics into account, we also rederive all our results for
$I_{40} = 4$. Future observations will show which value is closer to reality. 

\section{Fitting Models}

 The fitting model consists of three independent components: the cosmology,
the evolution of the GRB source population, and the intrinsic luminosity 
function (hereafter just luminosity function or LF).

 The cosmological model is not very important for the purpose of the present 
work as it affects only large redshifts 
while the main issue we are concerned with here is the source evolution at low redshifts. 
We adopted a flat vacuum-dominated cosmology ($\Omega_\Lambda = 0.7, \Omega _M = 0.3$) 
which is supported by recent data (see, e.g., Lukash, 2000).

 The evolution of the source population is the objective of our study. We 
tested four evolutionary functions. The first is a non-evolving population
(NE).
The second  is the star formation function, which is a reasonable hypothesis 
for
the evolution of GRB progenitors if they are collapsars. Porciani \& Madau (2001)
suggest three parameterized versions of the star formation rate which
are very similar at small redshifts, but differ at large redshifts
where the interpretation of the data is ambiguous due to the poorly known effects of dust 
absorption. Again, the evolution at large redshifts is beyond the scope of 
this work and we considered only one of these versions, namely:

\begin{equation}
R_{SF}(z) = \frac {0.15e^{3.4z}} {(e^{3.4z}+22)}  \quad
{\rm M_{\odot}} {\rm yr}^{-1} {\rm Mpc}^{-3}
\end{equation}

This expression describes a constant SF rate at large redshifts and 
corresponds to a comoving volume.

The two other evolution functions used correspond to neutron star merger
models. We obtained them by convolving the above SF 
rate with two different distributions for the delay between the formation of a 
binary system and the coalescence of its daughter neutron star binary.
The first delay distribution was taken from Lipunov et al. (1995), 
hearafter L95, and
the second from Portegies-Zwart \& Yungelson (1998), hereafter PZY98. 
These distributions are quite different from one another. L95 predicts a peak
at delays of 10 - 20 Myr and a long tail with a comparatively high 
probability of several Gyr delays. The distribution of PZY98 has a maximum
around 1 Gyr and lower probability at several Gyr.
  
The standard candle log $N$ - log $P$ 
distributions for these four models are shown in Figure 3.

 In addition to four fixed evolution models we tested different
slopes of the decline phase of the source population, modifying Eq. (1) as 

\begin{equation}
R_{SF}(z) \propto \frac {e^{1.086 ln(a+1)z}} {(e^{1.086ln(a+1)z}+ a)}  \quad
\end{equation}

where $a$ is a parameter describing the fall-off with redshift: (a+1) is 
the ratio of the comoving rates of GRB emission at  large $z$ and at $z=0$.
The expression coincides with (1) at $a=22$.

\centerline{\epsfxsize=10cm\epsfysize=10cm {\epsfbox{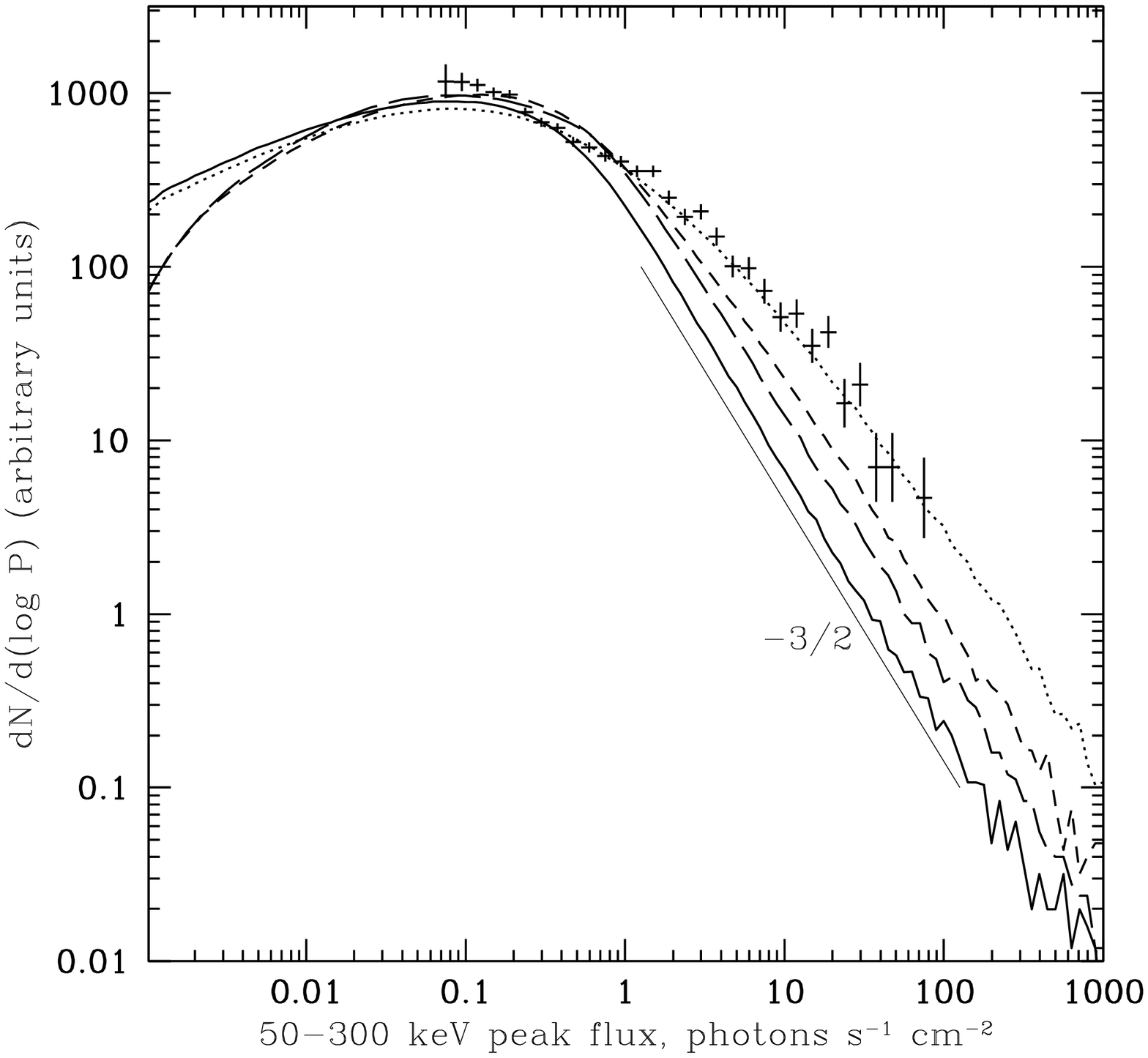}} }
\figcaption{
Standard candle log N - log P distributions for different models.
The standard candle brightness corresponds to a peak flux of 1 photon s$^{-1}$ cm$^{-2}$ at $z =1 $. Solid curve: SF model; 
dotted curve: NE model;
dashed curve: SF model convolved with L95 delay function; dash-dotted 
curve: SF model convolved with PZY98 delay function. The crosses represent the observed log N - log P distribution of 3255 long BATSE GRBs.  
 Theoretical curves are 
normalized to the same integral number of GRBs while data points are 
given in an arbitrary normalization.
}
\smallskip

 The third component of the model is the hypothetical luminosity function.
The data allow a wide choice with only 
two constraints: the width of the function, which must be at least 2.5 orders of magnitude 
(the luminosity range of GRBs with measured $z$), and the number
of intrinsically bright GRBs (see section 2). We chose a broken power law, which
proves sufficient freedom with a reasonably small 
number of free parameters:

  $dN/dP = C\cdot P^{\alpha -1}$ for $P_1<P<P_b$, $dN/dP = 
C_1\cdot P^{\alpha +\beta -1}$ for $P_b<P<P_2$ and $dN/dP = 0 $ beyond
the interval [$P_1,  P_2$]. The free parameters are
$\alpha$, $\beta$, $P_1$, $P_b$, and $C$, while $P_2$ was fixed to a 
value above the  maximum observed GRB brightness. 

We used the forward folding 
method when fitting GRB data,
i.e.,  the hypothetical brightness
distribution was convolved with the  efficiency matrix (1) and fitted to the
observed differential log N - log P distribution (crosses in
Figure 2) represented by 29 data points below $P = 50$ photons s$^{-1}$ cm$^{-2}$
In 9.1 years of BATSE and \it Ulysses \rm data, there were 15 GRBs 
brighter than this. We  treat the range $P>50$  photons s$^{-1}$ cm$^{-2}$ separately,
estimating the likelihood function of the fit for each peak flux range. For 
the main interval, this is  the standard $\chi^2$ probability function. For the tail
of the brightness distribution, the likelihood is the Poisson probability of
finding no more than 15 events apparently brighter than 
50 photons s$^{-1}$ cm$^{-2}$  assuming an average number
$A_{50}$ predicted by the model.  The final
likelihood finction is the product of these two factors.

\section{Results}

The best fit integral log N - log P distributions for the four models, SF, PZY98, L95, and NE
 are shown in figures 4 and 5. 

\centerline{\epsfxsize=9.5cm\epsfysize=9.5cm {\epsfbox{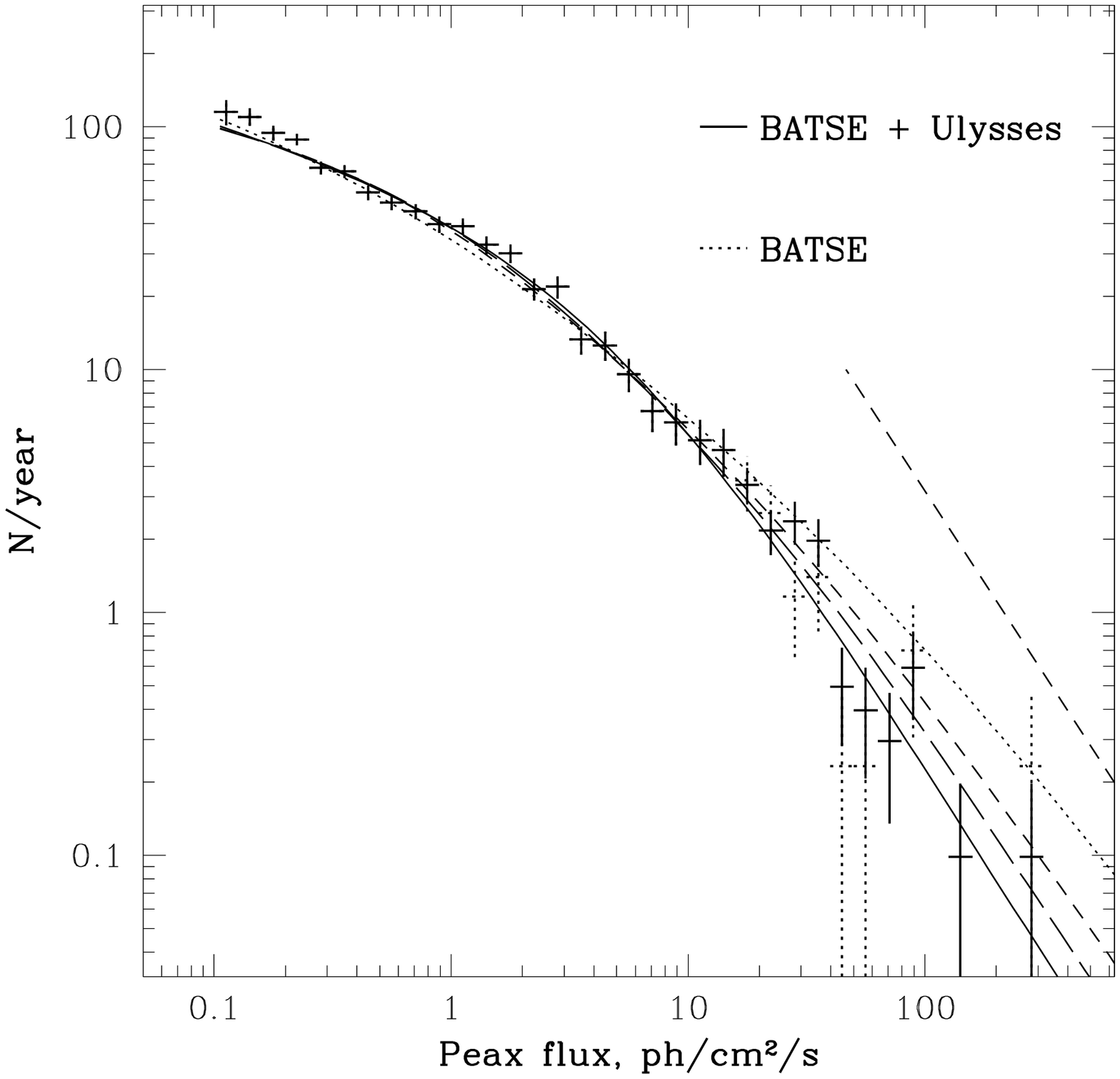}} }
\figcaption{
Comparison between the predictions of different evolutionary models and the data.
Solid curve: SF; dotted curve: NE; dashed curve: L95; long dashed curve: PZY98;
straight dashed line: the Euclidean -3/2 slope. 
Crosses are the observed data points, as in figure 2.
}


\centerline{\epsfxsize=9.5cm\epsfysize=9.5cm {\epsfbox{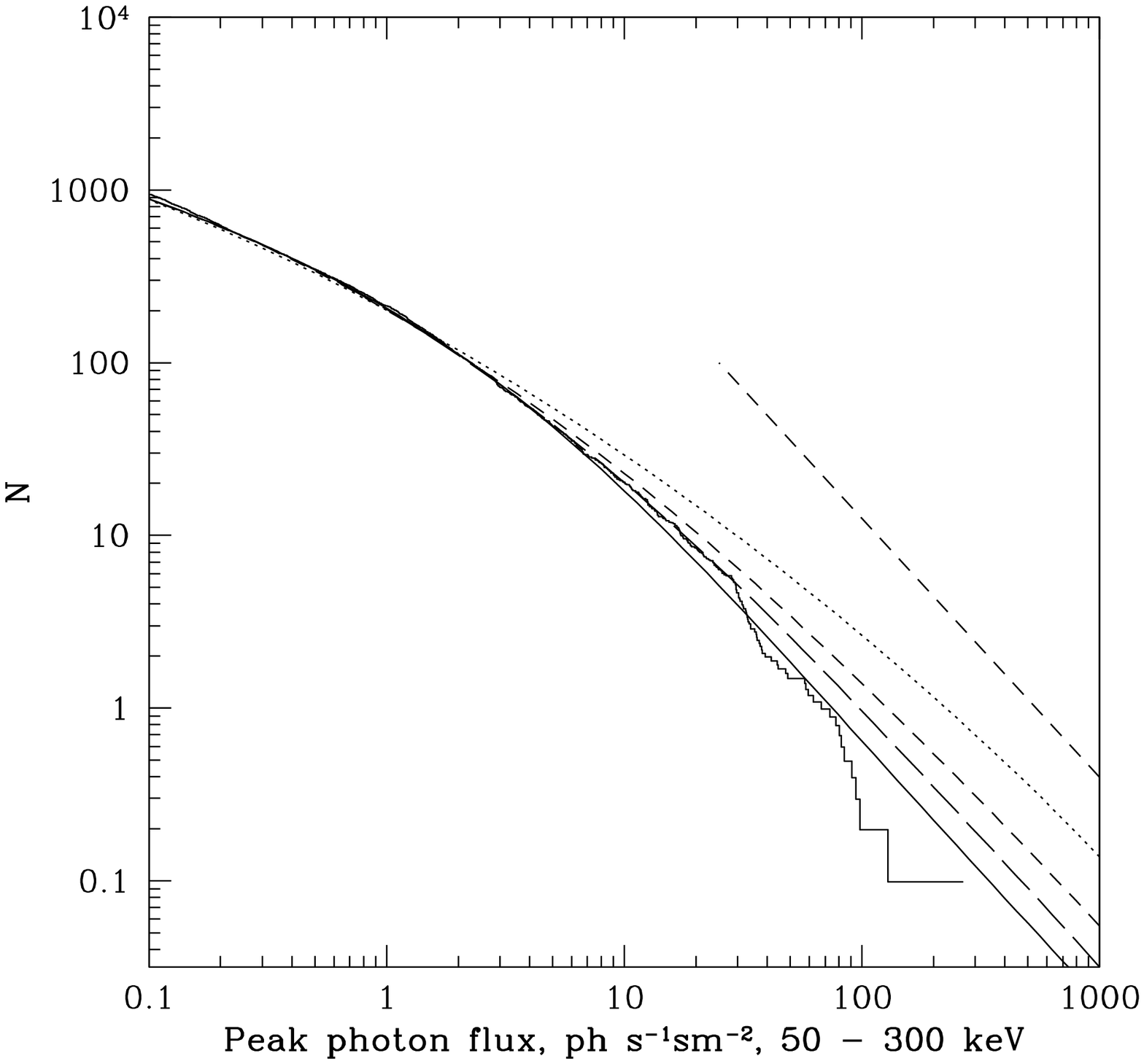}} }
\figcaption{
The same curves as in figure 4, but in integral form. 
Histogram - the integral peak flux distribution of \it Ulysses \rm /BATSE GRBs.
Solid curve: - SF model; dotted curve: NE; dashed curve:  L95; long dashed curve: PZY98. 
}
\smallskip

Adopting a rate of
intrinsically bright GRBs $I_{40} = 10$, their likelihoods are 0.034,  1.9 $\cdot 10^{-3}$, 
0.45 $\cdot 10^{-5}$ and 2.2 $\cdot 10^{-16}$ respectively.
If we overestimate the rate of intrinsically strong GRBs by a
factor 2.5 (assuming that 4 of the observed events with $P_I > 40$ are a fluctuation, so  
that $I_{40} = 4$, then the likelihoods are 0.40, 0.015,
0.95$\cdot 10^{-3}$, 2.65$\cdot 10^{-9}$ respectively.

 Note that the rejection of the NE model ($10^{-16}$) is now much stronger 
than in STS 
($10^{-4}$). This improvement is partially due to the better statistics  
of the joint BATSE-\it Ulysses \rm  sample, but mainly due to the fact that STS used too low an
estimate for $I_{40}$: 3 events per year, inferred from the conservative 
assumption that $F_s \sim 0.5$ (compared to the present estimate $F_s \sim 0.1$
obtained using \it Ulysses \rm  data).

The sharp break in figure 5 around $P_a \sim 30$ photons s$^{-1}$ cm$^{-2}$
is statistical in origin (note the corresponding deviation of the two data 
points in figure 4). This fluctuation is mainly due to the
\it Ulysses \rm  sample; it is evident in Atteia et al. (1999).

\centerline{\epsfxsize=10cm\epsfysize=10cm {\epsfbox{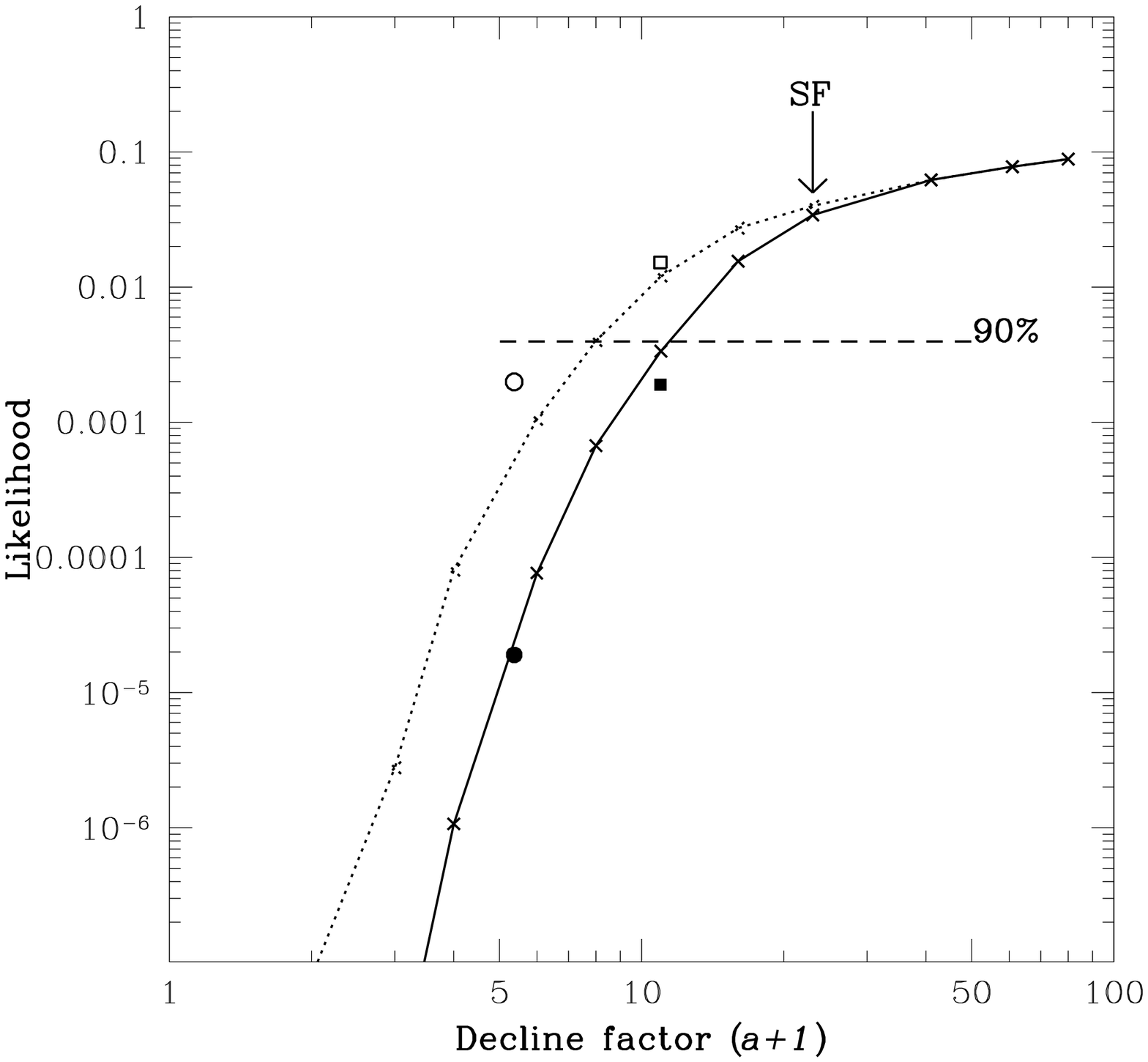}} }
\figcaption{ The likelihood versus the fall-off factor ($a+1$) where $a$
is the parameter in equqtion 2.
 The arrow shows the result for the SF model; the 90\% confidence 
limit (dashed horizontal line) is 
given with respect to this model.
}
\smallskip

 Figure 6 shows the likelihood factor for the parametrized source evolution model
(equation 2) versus the fall-off factor $a + 1$. The results for NS merger models 
are also shown; the ordinate for these models is just the ratio
of the maximal NS merging rate (at $z \sim 2$) to that at $z=0$.
The likelihood curve has no turover at large $a$ because our
luminosity fuction has only a lower limit constraint at its bright end.
It is interesting that the curve still displays
a considerable increase (by a factor of 2.6) from $a$=22, which corresponds to the
SF curve, to $a=80$, i.e., the data are better fit by a GRB progenitor fall-off which is faster than 
the SF rate. This could be a natural consequence if the progenitors are supermassive stars whose population 
can decline faster than the total SF.  However, this indication is statistically weak and could also result, for example,
from the same fluctuation in the \it Ulysses \rm  data which produces a break in the log N
- log P curve around 20 - 30  photons s$^{-1}$ cm$^{-2}$, as discussed 
above.    

More details of the fits  are presented in Table 1,
where $\chi^2$ and the predictions for $A_{50}$ are given 
separately. We do not present the best fit parameters for the broken power law luminosity function 
because they are consistent with the results of STS where
this issue was studied in detail.
 
Table 1, as well as figures 4 and 5, clearly demonstrate that models with 
an insufficiently steep evolutionary decline of GRB progenitors  predict an excess of 
apparently strong GRBs with respect to the observations.  

\smallskip
\begin{tabular}{ccccc}
\hline
Model& lkh& $\chi^2$ & $A_{50}$ &lkh($A_{50}$)\\
\hline
NE&2.2$\cdot 10^{-16}$&83&49&1.28$\cdot 10^{-8}$\\
SF$_{22}$&0.034&31&18.6&0.24\\
L85&1.9$\cdot 10^-{5}$&37&33&2.1$\cdot 10^{-4}$\\
PZY98&1.9$\cdot 10^{-3}$&32&26&1.4$\cdot 10^{-2}$\\
SF$_{80}$&0.088&31 &15 &0.57\\
SF$_{40}$&0.062&31 &16 &0.47\\
SF$_{10}$&0.36$\cdot 10^{-2}$&31&25&2.2$\cdot 10^{-2}$\\
SF$_{5}$&0.76$\cdot 10^{-4}$&38&30&1.9$\cdot 10^{-3}$\\
\hline
\end{tabular}
\vskip0.1cm
{\small{\bf Table1} The maximum likelihood results for various models.
The second column (lkh) gives the final likelihood factor; the third column, the
$\chi^2$ value (at 24 degrees of freedom); the fourth, the predicted $A_{50}$ (the observed $A_{50}$ is 15);
the fifth, the probability of observing $A_{50}$ less than 16 for its predicted 
value. The subscripts in the first column correspond to the value of $a$ 
(see equation 2).  SF$_{22}$ in row 2 corresponds to the measured star 
formation curve.
}  
\smallskip

 If we use the Bayesian approach, treating the ratio of likelihoods 
as the relative probabilities of different models, then the rejection factors
for NS models relative to SF are 0.055 for PZY98 and 
1.3$\cdot 10^{-3}$ for L95. If we adopt the estimate $I_{40} = 4$ 
then the constraints relax to 0.37 and 0.024 respectively, i.e. the PZY98
model is consistent with the data. Note however that the choice of $I_{40} = 4$
corresponds to a less than 0.1 probablity fluctuation in the number of intrinsically 
bright GRBs with measured redshifts. 

 The minimal fall-off factor allowed by the data at 90\% confidence level is
$a+1 \sim 12$ or $a+1 \sim 8$ for $I_{40} = 4$.

\section{Conclusions}

 The joint BATSE - \it Ulysses \rm  data confirm a sharp decline in the GRB source population between
$z \sim 2$ and 
the present epoch. Although it is consistent with that of star 
formation, a faster decline is slightly preferable, albeit at a 
statistically insignificant level ($\sim 1\sigma$).
The two models of binary system evolution leading to a final NS merger
are well beyond the 90\% confidence limit, except for the $I_{40} = 4$
case,  which is based on the assumption of a large fluctuation in the observed number of  intrinsically
bright GRBs.
   Note that while the statistics of bright GRBs will improve slowly,  the redshift statistics
can improve much faster, so that a more reliable estimate of $I_{40}$ 
may be available relatively soon.

The joint BATSE/\it Ulysses \rm  data present a new challenge to the neuton star binary model
as an explanation of the source of long GRBs. Together with the results of afterglow studies it makes it very 
improbable. The only way to save the NS model is to show that the 
typical lifetimes of such systems is short.  If very few survive longer than 1 Gyr,
this will fit the log N - log P distribution, and if many merge in a few  Myr,
this will explain the locations of the observed afterglows in the star forming
regions of their host galaxies. Such a possibility has been studied in the recent 
work of Belczynski, Bulik \& Rudak (2001) where it is shown that this could occur 
in some  binary evolution models due to common 
envelope events producing very tight NS systems.  
 Finally it should be pointed out again that our constraints
 refer only to the class of long GRBs, while the NS binary model 
 is probably able to explain the origin of short bursts. 


We thank Ya.Tikhomirova for assistance.
This research made use of data obtained through the HEASARC Online Service
provided by NASA/GSFC.
It was supported by the Swedish Natural Science Research Council,
the Royal Swedish Academy of Science, the Wennergren Foundation for 
Scientific Research and the Russian Fondation for Basic Research (grant 00-02-16135).
KH is grateful for \it Ulysses \rm support under JPL Contract 958056.


\end{document}